\documentstyle[amssymb,aps,prl,epsfig]{revtex}
\begin{document}
\draft
\twocolumn[\hsize\textwidth\columnwidth\hsize\csname %
@twocolumnfalse\endcsname

\title{Spectral functions and pseudogap in the $t$-$J$ model}
\author{P. Prelov\v sek and A. Ram\v sak}
\address{Faculty of Mathematics and
Physics, University of Ljubljana, 1000 Ljubljana, Slovenia }
\address{J. Stefan Institute, University of Ljubljana, 1001 Ljubljana,
Slovenia }
\date{\today}
\maketitle 
\begin{abstract}
\widetext  

We calculate spectral functions within the $t$-$J$ model as relevant
to cuprates in the regime from low to optimum doping. On the basis of
equations of motion for projected operators an effective spin-fermion
coupling is derived. The self energy due to short-wavelength
transverse spin fluctuations is shown to lead to a modified
selfconsistent Born approximation, which can explain strong asymmetry
between hole and electron quasiparticles. The coupling to
long-wavelength longitudinal spin fluctuations governs the
low-frequency behavior and results in a pseudogap, which at
low doping effectively truncates the Fermi surface, in particular near
the $(\pi,0)$ point in the Brillouin zone.

\end{abstract}
 ]
\narrowtext
In recent years underdoped cuprates are in the center of experimental
and theoretical investigation, offering possibly a clue to the
understanding of anomalous normal state properties and the mechanism
of superconductivity in these compounds. Here we concentrate on some
experimental facts revealing the nature of quasiparticles (QP) and
the pseudogap. Several quantities, in particular the uniform
susceptibility, the Hall constant, the specific heat, show the (large)
pseudogap scale $T^*$ \cite{imad}, consistent with the angle resolved
photoemission (ARPES) revealing a hump at $\sim 100$~meV observed in
Bi$_2$Sr$_2$CaCu$_2$O$_{2+\delta}$ (BSCCO) near the $(\pi,0)$ momentum
\cite{mars}. At the same time QP dispersing through the Fermi surface
(FS) are resolved by ARPES in BSCCO only in parts of the large FS, in
particular along the nodal $(0,0)$-$(\pi,\pi)$ direction, indicating
that the rest of the large FS is either fully or effectively
gaped. On approaching the optimal doping the $T^*$ scale merges with
$T_c$ and the large FS becomes well defined. All these phenomena are
naturally associated with the antiferromagnetic (AFM) short-range
order (SRO) in cuprates, since the scale $T^*$ approaches the AFM
exchange $J$ in undoped materials. This is less clear for the lower
spin-gap scale $T_{sg}$ (not the subject here), not found in all
cuprates.

While these facts allow for a qualitatively consistent scenario, the
consensus on necessary prerequisites and moreover a satisfactory
theoretical analysis are still missing. The renormalization group
studies of the Hubbard model \cite{zanc} indicate on the
breakdown of the standard Fermi liquid and on the truncation of the
FS.  That such features also emerge from prototype models of
correlated electrons has been confirmed in numerical studies of
spectral functions in the Hubbard \cite{preu} and in the $t$-$J$ model
\cite{jpspec,jprev}, which both show the appearance of the pseudogap
at low doping. Some aspects of the pseudogap have been found in the
spin-fermion models \cite{pine} and studied phenomenologically in
the Hubbard model \cite{dahm}.

Our aim is to capture these features within an analytical treatment of
a single band model. In
the following we show that an effective spin-fermion model can be
derived via equations of motion (EQM) and dividing the coupling into
short and long-wavelength spin fluctuations an approximation for the
electron self energy can be found.

We study the planar $t$-$J$ model
\begin{equation}
H=-\sum_{i,j,s}t_{ij} \tilde{c}^\dagger_{js}\tilde{c}_{is}
+J\sum_{\langle ij\rangle}({\bf S}_i\cdot {\bf S}_j-\frac{1}{4}
n_in_j) , \label{eq1}
\end{equation}
where we take into account possible longer range hopping, i.e.,
besides $t_{ij}=t$ for n.n. hopping also $t_{ij}=t'$ for n.n.n.
neighbors on a square lattice. We evaluate the single-particle propagator
in this model explicitly taking into account that fermionic operators
are projected ones not allowing for the double occupancy of sites, e.g.,
$\tilde{c}^\dagger_{is}= (1-n_{i,-s}) c^\dagger_{is}$.

We use EQM directly for projected operators \cite{prel}
and represent them in variables appropriate for a paramagnetic
metallic state with $\langle {\bf S}_i \rangle =0$ and electron
concentration $\langle n_i
\rangle = c_e=1-c_h$,
\begin{eqnarray}
  [\tilde c_{{\bf k} s},H]&=& [(1-\frac {c_e}{2}) \epsilon^0_{\bf k} -
  J c_e]\tilde c_{{\bf k} s} + \frac{1}{\sqrt{N}} \sum_{\bf q} \bigl (
  2J \gamma_{\bf q} + \epsilon^0_{{\bf k}-{\bf q}} \bigr ) \nonumber\\
  &&\bigl[ s S^z_{\bf q} \tilde c_{{\bf k}-{\bf q},s} + S^{\mp}_{\bf
    q} \tilde c_{{\bf k}-{\bf q},-s} - \frac{1}{2} \tilde n_{\bf q}
  \tilde c_{{\bf k}-{\bf q}, s}\bigr],
\label{eq2}
\end{eqnarray}
where $\epsilon^0_{\bf k}=-4t\gamma_{\bf k}-4t'\gamma'_{\bf k}$ is the
bare band energy on a square lattice and $\gamma_{\bf k}=(\cos
k_x+\cos k_y)/2$, $\gamma_{\bf k}'=\cos k_x \cos k_y$.

EQM for $\tilde c_{{\bf k} s}$ can be used to construct approximations
for the electron propagator $G({\bf k},\omega)$  \cite{prel,plak},
which can be represented as
\begin{equation}
  G({\bf k},\omega)= \frac{\alpha}{\omega+\mu -\zeta_{\bf k} - \Sigma({\bf
      k},\omega) }, \label{eq3}
\end{equation}
where the renormalization $\alpha= (1+c_h)/2$ is a consequence of the
projected basis, $\mu$ is the chemical potential,  
and $\zeta_{\bf k}$ is the 'free' propagation term
emerging from the EQM,
\begin{equation}
  \zeta_{\bf k}= \frac{1}{\alpha} \langle \{[\tilde c_{{\bf k} s},H],
  \tilde c^{\dagger}_{{\bf k} s}\}_+\rangle -\bar \zeta
= -4 \eta_1 t \gamma_{\bf k}
  -4 \eta_2 t' \gamma'_{\bf k}, \label{eq4}
\end{equation}
where $\eta_j = \alpha + \langle {\bf S}_0 \!\cdot\! {\bf S}_j
\rangle/\alpha$ and  $\bar \zeta$ is
a constant. The central
quantity for further consideration is the self energy $\Sigma({\bf
k},\omega) = \langle\!\langle C_{{\bf k}s};C^+_{{\bf k}s}
\rangle\!\rangle_\omega^{irr} /
\alpha $, where $iC_{{\bf k}s}=[\tilde c_{{\bf k} s},H]-\zeta_{\bf k}
\tilde c_{{\bf k} s}$, and only the 'irreducible' part of the correlation
function should be taken into account in the evaluation of $\Sigma$.
In finding an approximation for $\Sigma$ we assume that we are dealing
with the paramagnet with pronounced AFM SRO with the dominant wave
vector ${\bf Q} =(\pi,\pi)$ and the AFM correlation length $\xi>1 $
with corresponding $\kappa =1/\xi$. We first note that EQM,
Eq.~(\ref{eq2}), naturally indicate on an effective coupling between
fermions and spin degrees. However, the role of short-range and
longer-range spin fluctuations is quite different.

In an undoped system AFM the spectral function of an added hole is
quite well described within the selfconsistent Born approximation
(SCBA) \cite{kane}, where the strong hole-magnon coupling induced by
the hopping $t$-term leads to a broad background representing the
incoherent hopping and a narrow QP dispersion governed predominantly
by $J$. If we assume as a starting point an undoped N\'eel state as
well as $J < t$  EQM (\ref{eq2}) directly
reproduce the coupling equivalent to the holon-spin coupling within
the SCBA. Note that within the N\'eel (Ising) state we have
$\eta_1=0$, $\eta_2=1$, $S^z_i=\pm 1/2$, and a nontrivial coupling
comes from transverse $S^\mp_{\bf q}$ which can be represented via
magnon excitations.  Therefore by performing the decoupling of fermion
and spin degrees in $\Sigma$ and by using the identity $\tilde
c_{js}=\tilde c_{j,-s} S^\mp_j$, we recover the standard SCBA
equations.

Our EQM formalism thus naturally leads to the SCBA in an undoped
system. Still we note that in an isotropic AFM $\eta_1\ne 0$ (but
$|\eta_1| \ll 1$) which slightly modificate SCBA results. Since
the SCBA accounts well for properties of a single QP in an AFM, we are not
trying here to improve it. Within our approach we
generalize the equations for finite doping $c_h>0$ where we have 
electron-like QP above the Fermi energy ($\omega>0$). In 2D the AFM
long-range order is absent due to $T>0$ and $c_h>0$, still spin
fluctuations are magnon-like, i.e., propagating and transverse to the
local AFM SRO, with a dispersion $\omega_{\bf q}$ for $q>\kappa$ and
$\tilde q>\kappa$ where $\tilde {\bf q}= {\bf q}-{\bf Q}$. Hence the
paramagnon contribution to the self energy can be written as
\begin{eqnarray}
  &&\Sigma_{\rm pm}({\bf k},\omega)= \frac{16t^2}{N} \sum_{q,
\tilde q> \kappa}
  (u_{\bf q} \gamma_{{\bf k}-{\bf q}}+v_{{\bf q}} \gamma_{\bf k})^2
  \nonumber \\ &&[ G^-({\bf k}-{\bf q},\omega+\omega_{\bf q}) +
  G^+({\bf k}+{\bf q},\omega-\omega_{\bf q})], \label{eq5}
\end{eqnarray}
where $(u_{\bf q},v_{\bf q})=(1,-{\rm sign}(\gamma_{\bf q}))\sqrt{(2J
\pm\omega_{\bf q})/2\omega_{\bf q}}$ and $G^\pm$ refer to the Green's
functions corresponding to electron ($\omega>0$) and hole ($\omega<0$)
QP excitations, respectively. So far equations are written for $T=0$,
however in $\Sigma_{\rm pm}$ the role of finite but low $T>0$ is not
pronounced. Note that analogous to the SCBA $t'$ does not enter
directly the coupling but remains in the 'free' propagation term
$\zeta_{\bf k}$.  Here we stress two features of our generalized SCBA:
a) we are dealing with a strong coupling theory due to $t >
\omega_{\bf q}$ hence a selfconsistent calculation of $\Sigma$ is required, b)
resulting spectral functions $A({\bf k},\omega)$ are very asymmetric
with respect to $\omega=0$, since $G^+$ has less weight and
consequently the scattering of electron QP is less pronounced.

We are dealing with a paramagnet, therefore it is essential to consider
also the coupling to longitudinal spin fluctuations. Note that the EQM
(\ref{eq2}) naturally introduce a coupling between fermion and spin
operators which is isotropic in the spin space as appropriate in a
paramagnetic state. In fact analogous form as in Eq.~(\ref{eq2}) would
emerge also from a spin-fermion Hamiltonian with the coupling
parameter $m_{\bf kq}= 2J \gamma_{\bf q} +\epsilon^0_{{\bf k}-{\bf
q}}$. The effective Hamiltonian should be hermitian, i.e., the
coupling should satisfy the condition $\tilde m_{{\bf k},{\bf q}} =
\tilde m_{{\bf k}-{\bf q},-{\bf q}}$, therefore we use further on the
symmetrized $\tilde m_{\bf kq}= 2J
\gamma_{\bf q}+ \frac{1}{2} (\epsilon^0_{{\bf k}-{\bf q}}+\epsilon^0_{\bf k})$.

Fermions and longitudinal spin fluctuations with $\tilde q<\kappa$
appear to be quite uncoupled, therefore we express the longitudinal
contribution as in Refs.~\cite{prel,plak},
\begin{eqnarray}
\Sigma_{\rm lf}({\bf k},\omega) &=&\frac{1}{\alpha N} \sum_{\bf q}
\tilde m^2_{\bf k q}
\int \int \frac{d\omega_1 d\omega_2}{\pi} g(\omega_1,\omega_2) \nonumber\\
&&\frac{A^0({{\bf k}-{\bf q}},\omega_1) \chi''({\bf q},\omega_2)}
{\omega-\omega_1-\omega_2 }, \label{eq7}
\end{eqnarray}
where $\chi({\bf q},\omega)$ is the dynamical spin susceptibility,
$A^0({\bf k},\omega)=-(\alpha/\pi){\rm Im}(\omega +\mu -\zeta_{\bf k} -
\Sigma_{\rm pm})^{-1}$ and
$g(\omega_1,\omega_2)={\rm th}(\omega_1/2T)+{\rm cth}(\omega_2/2T)$.
In $\Sigma_{\rm lf}$ only the part corresponding to irreducible
diagrams should enter, so there are restrictions on proper
decoupling. We are mostly dealing with the situation with a pronounced
AFM SRO where longitudinal spin fluctuations are slow, with a
characteristic frequency $\omega_\kappa \ll J$ which is the case of a
quasistatic $\chi({\bf q},\omega)$.  Therefore we in
Eq.~(\ref{eq7}) as the simplest approximation  insert the
unrenormalized $A^0({\bf k},\omega)$, i.e., the spectral function
without a self-consistent consideration of $\Sigma_{\rm lf}$ but with
$\Sigma_{\rm pm}$ fully taken into account.  Such an approximation has
been introduced in the theory of a pseudogap in charge density wave
systems
\cite{lee}, used also in related works analyzing the role of spin
fluctuations \cite{kamp},\cite{chub}, and recently more extensively
examined in Ref.\cite{mill}.

So far we do not have a corresponding theory for the spin response at
$c_h>0$ and $T>0$, so $\chi({\bf k},\omega)$ is assumed as a
phenomenological input, bound by the sum rule
\begin{equation}
  {1\over N} \sum_{\bf q} \int_0^{\infty} {\rm
    cth}(\frac{\beta\omega}{2}) \chi''({\bf q},\omega) d \omega =
  \frac{\pi} {4} (1-c_h). \label{eq8}
\end{equation}
The response should qualitatively correspond to a paramagnet close to
the AFM instability, so we assume the form
\begin{equation}
  \chi''({\bf q},\omega) \propto\frac{ \phi(\omega,T)}{
(\tilde q^2 +\kappa^2) (\omega^2+\omega_\kappa^2)}, \label{eq9}
\end{equation}
where $\phi(\omega,T) \propto \omega$ would be appropriate for a
nearly AFM Fermi liquid \cite{pine,chub} or an undoped AFM in 2D at
any $T>0$. On the other hand, in cuprates at intermediate doping more
consistent with model results for $T>0$ seems to be the marginal
Fermi liquid behavior  with $\phi(\omega,T) \propto {\rm th}(\omega/2T)$
\cite{varm,imad,jprev}.

Eqs.~(\ref{eq5},\ref{eq7}) for $\Sigma=\Sigma_{\rm pm}+\Sigma_{\rm
lf}$ represent the selfconsistent set of equations for $G$.
Parameters $\kappa, \eta_1, \eta_2$ are mainly dependent on $c_h$ and
are known from model calculations \cite{bonca89,imad}. At $T=0$ and
given $c_h$ we determine $\mu$ such that the density of states
${\cal N}(\omega)=(2/N)\sum_{\bf k} A({\bf k},\omega)$ integrated for
$\omega<0$ reproduces $c_e$. At the same time FS is 
given by the relation $\zeta_{{\bf k}_F} + \Sigma'({{\bf
k}_F},0)= \mu$. Full numerical analysis of selfconsistent equations will
be presented elsewhere. Here we concentrate on some key aspects of the
theory.  One is that $\Sigma_{\rm pm}$ allows for a meaningful
behavior in the limit $c_h \to 0$, which has been the deficiency of
most phenomenological theories so far. In this limit our results for
$A({\bf k},\omega)$ at $T \sim 0$ are essentially equivalent to
results within the SCBA approach \cite{kane} .  For $c_h \agt 0$ we
have a finite contribution from electron QP $A^0({\bf k},\omega>0)$
and the corresponding QP density evolves as $\propto c_h$.

We choose further on parameters $J=0.3t, t'=-0.2t$ and $\kappa=2
\sqrt{c_h}$. In Fig.~1 we present typical
results for $A({\bf k},\omega)$ along the $(0,0)-(\pi,\pi)$
direction. Since at low doping $\eta_2 \sim 0.9$ main doping
dependence arises from $\eta_1(c_h)$, which varies from $\eta_1
\sim -0.18$ at $c_h \to 0$ to $\eta_1 \sim \alpha$ for large
doping. In calculation we use 
$\eta_i$  close to values emerging from spin       
correlations found numerically \cite{bonca89}.

As presented in
Fig.~1 $\Sigma_{\rm pm}$ leads to a strong 
damping of the hole QP and quite incoherent
momentum-independent spectrum $A({\bf k},\omega)$ for $\omega \ll -J$
which qualitatively reproduces ARPES and numerical results
\cite{jprev}.
Electron QP (at $\omega>0$) are in general very
different,i.e., with much weaker damping arising only from
$\Sigma_{\rm pm}$. We should note that at given $\mu$, $c_e$
calculated from the density
of states does not in general coincide with
the one 
\vskip -.5cm
\begin{figure}[htb]
\center{\epsfig{file=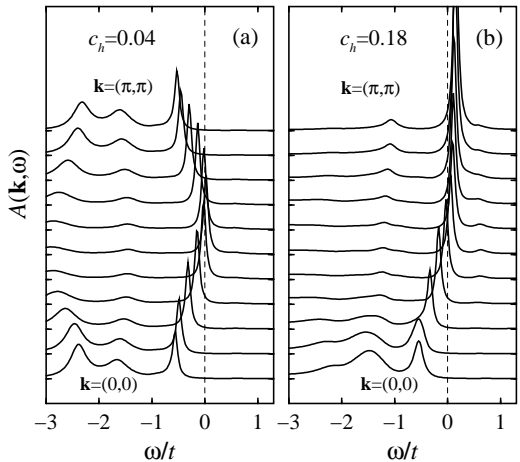,height=70mm,angle=-90}}
\vskip .2 cm
\caption{$A({\bf k},\omega)$ along the
$(0,0)-(\pi,\pi)$ direction for $J=0.3t$ and $t'=-0.2t$.
(a) $c_h= 0.04$, $\eta_1=0.02$ and (b) $c_h= 0.18$, $\eta_1=0.2$.}
\end{figure}
evaluated 
from the FS volume, $\tilde c_e=V_{\rm FS}/V_0$.
Nevertheless, apart from the fact that within the $t$-$J$ model
the validity of the Luttinger theorem is anyhow under question
\cite{putt}, in the regimes of large FS both quantities appear to be
quite close.

Results for a characteristic development of the FS with $c_h$ are
shown in Fig.~2. At $c_h<c_h^* \sim 0.08$ solutions are consistent
with a small pocket-like FS, whereby this behavior is enhanced by
$t'<0$ as realized in other model studies \cite{tpr}. On increasing
doping $c_h>c_h^*$ the FS rather abruptly changes from
small into a large one. The smallness of $c_h^*$ has the origin in
quite weak dispersion dominated by $J$ and $t'$ at $c_h
\to 0$ which is overshadowed by much larger $\zeta_{\bf k}$ at moderate
doping, where the FS is large and its shape is controlled by $t'/t$.
Nevertheless, the relevance of obtained FS should be considered in
connection with a coexistent pseudogap discussed further on.

The position of the FS is mainly determined by $\zeta_{\bf k}$ and
$\Sigma_{\rm pm}$, while in this respect $\Sigma_{\rm lf}$ is less
crucial.  Results for $\Sigma_{\rm pm}'({\bf k},0)$ can be well
parameterized in the form obtained within the SCBA for a single hole
\cite{kane}.  Similarly one can present also the full effective QP
dispersion, $\epsilon_{\bf k}^{\rm ef}= \zeta_{\bf k} +
\Sigma'({\bf k},0)-\mu$, and the QP residue $Z_{\bf k}$.
The simplest approximation to discuss the pseudogap is the
quasi-static approximation which is meaningful for $\omega_\kappa \ll
t$. Assuming also $\kappa \ll 1$ simplifying $g \chi''({\bf q},\omega)
\sim \pi\delta({\bf q}-{\bf Q}) \delta(\omega)/4$, as well as
a single-pole form $A^0({\bf k},\omega)= \alpha Z_{\bf k} \delta
(\omega - \epsilon^{\rm ef}_{\bf k})$ near the FS, we obtain from
Eqs.~(\ref{eq3},\ref{eq7})
\begin{equation}
G({\bf k},\omega)= \frac{\alpha Z_{\bf k} (\omega -
\epsilon^{\rm ef}_{{\bf k}-{\bf Q}})} {(\omega - \epsilon^{\rm ef}_{{\bf
k}-{\bf Q}})(\omega - \epsilon^{\rm ef}_{\bf k}) - \Delta^2_{{\bf k Q}} },
\label{eq11}
\end{equation}
where the gap function is given by $\Delta^2_{{\bf k Q}} = Z_{\bf k}
Z_{{\bf k}-{\bf Q}} \tilde m^2_{\bf kQ}/4$.  From resulting branches
$E^\pm$ it is evident that a gap opens on the AFM zone boundary, so
that the relevant pseudogap energy is $\Delta^{PG}_{\bf
k}=|\Delta_{\bf k_{AFM}}| \sim Z_{\bf k} |2J - 4 t' {\rm cos}^2
k_x|/2$.  For $t'<0$ the gap is largest at $(\pi,0)$ point, as
observed in experiments \cite{mars}. 
Since within the same ${\bf k}$
region the QP $\,\,$ dispersion is $\,\,$ also smallest $\,$ the  
\vskip -.2 cm     
\begin{figure}[htb]
\center{\epsfig{file=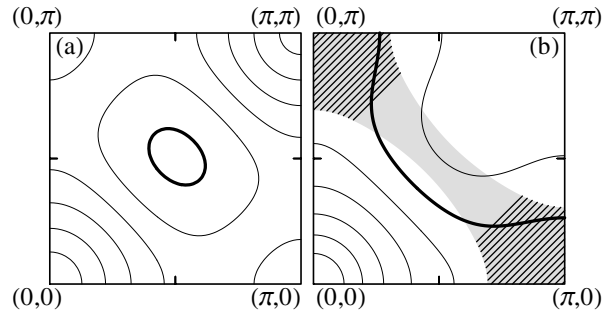,height=45mm,angle=-0}}
\caption{Fermi surface corresponding to results presented in Fig.~1.
(a) Small FS, $c_h=0.04$ and (b) large FS, $c_h=0.18$.  Contour lines
represent QP energy levels in increments $0.1 t$. 
The region with a developed pseudogap ($w/\Delta<1$) is line-shaded
while the
grey-shaded region represents the region where the pseudogap is
smeared out ($w/\Delta>1$).}
\end{figure}
$\!\!\!\!\!\!$effect 
is even more
pronounced. If $\omega = 0$ is in the regime of the gap, then
naturally we are dealing (in this approximation) with a truncated FS.
For parameters as above we present in Fig.~2(b) the pseudogap region
where the states and FS near the $(\pi,0)$ momentum are strongly
suppressed.

Going beyond the quasi-static and $\kappa \sim 0$ treatment one can
discuss also the QP states within the pseudogap. To study the general
structure of the SF in this region it is enough to follow the
development with $\epsilon=\epsilon^{\rm ef}_{\bf k}$ crossing the
pseudogap perpendicular to the AFM zone boundary. It is essential to
take into account $\kappa >0$ so that the averaging over $\tilde {\bf
q}$ leads to an effective smearing of the delta-function $A^0_{{\bf
k}-{\bf Q}}$ into a broader $\bar A(\epsilon,\omega)$. So we have
qualitatively to deal (at $T=0$) with the self energy
\begin{equation}
  \Sigma''(\epsilon,\omega) \propto \int_0^\omega
  \chi''(\omega-\omega') \bar A(\epsilon,\omega') d\omega',
  \label{eq12}
\end{equation}
where the simplest assumption for $\bar A(\epsilon,\omega)$ is a
Lorentzian with the width $w= v_{{\bf k}_F}\kappa$. Analogous
equations have been already studied in Ref.~\cite{kamp} and lead to
the pseudogap of the order of $\Delta=\Delta_{\bf k}^{PG}$, pronounced
in QP spectra and clearly in ${\cal N}(\omega)$. Results for the case
with the gap centered at $\omega=0$ are shown in Fig.~3, where the
depletion is most evident for $w \ll \Delta$, while the pseudogap
fills up for $w > \Delta$.
\vskip -1.5 cm
\begin{figure}[htb]
\center{\epsfig{file=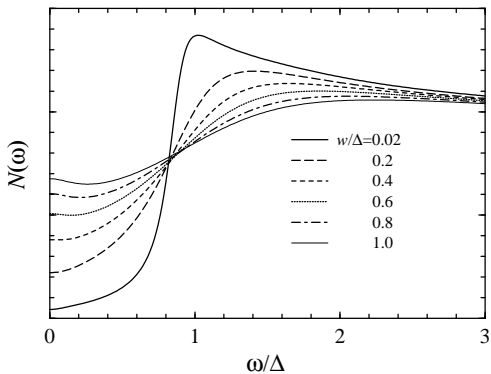,height=78mm,angle=-90}}
\vskip -.4 cm
\caption{Density of states ${\cal N}(\omega)$ as a function of $\omega/
\Delta$ for $\omega_{\kappa}=0.2 \Delta$. }
\end{figure}
Looking at corresponding spectral functions directly, we notice that
for a developed pseudogap with $w\ll \Delta$ (in Fig.~2(b) line-shaded)
there are still QP crossing the FS, although with small 
$Z_{\bf k} \ll 1$, while their velocity is not
diminished. On the other hand if $w>\Delta$ the pseudogap is smeared
out and consequently not effective, hence the FS is fully
recovered. At intermediate doping this typically happens near the zone
diagonal as shown in Fig.~2(b) (grey shaded).

In conclusion, we have presented a theory for the spectral functions
within the $t$-$J$ model where the double-occupancy constraint is
taken explicitly into account and used to derive an effective
spin-fermion coupling. The coupling to transverse AFM paramagnons is
strong, nevertheless it can be well treated within a generalized
SCBA. On the other hand, the coupling to longitudinal AFM
fluctuations, $\tilde m_{\bf kq}$, is moderate near FS for low doping
and leads to a pseudogap, fully developed near the $(\pi,0)$
point. The pseudogap is not in contradiction with the existence
of a large FS, and should show up in integrated photoemission and
ARPES results as well as in the uniform susceptibility and in the
specific heat. More elaborate analysis of proposed theory will be
presented elsewhere.

One of authors (P.P.) wishes to thank T.M. Rice, G. Khaliullin and M.
Imada for helpful discussions, and acknowledges the support of the
Japan Society for the Promotion of Science, and as well the Institute
for Theoretical Physics, ETH Z\"urich, Institute for Materials
Research, Sendai, and ISSP, University of Tokyo, Kashiwa, where part
of this work has been done.


\vskip -.3 cm
\begin{references}
\vspace{-1.3cm}        
\bibitem{imad} for a review see, e.g., M. Imada, A. Fujimori, and Y.
  Tokura, Rev. Mod. Phys. {\bf 70}, 1039 (1998).

\bibitem{mars} D.S. Marshall {\it et al.}, Phys. Rev. Lett. {\bf 76},
  4841 (1996); H. Ding {\it et al.}, Nature {\bf 382}, 51 (1996).

\bibitem{zanc} D.  Zanchi and H.J. Schulz, Europhys.  Lett. {\bf 44},
  235 (1997); N.  Furukawa, T.M. Rice, and M.  Salmhofer, Phys. Rev.
  Lett. {\bf 81}, 3195 (1998)

\bibitem{preu} R. Preuss, W. Hanke, C. Gr\" ober, and
  H.G. Evertz, Phys. Rev.  Lett. {\bf 79}, 1122 (1997).

\bibitem{jpspec} J. Jakli\v c and P.  Prelov\v sek, Phys. Rev. B {\bf
    55}, R7307 (1997); P. Prelov\v sek, J. Jakli\v c, and K. Bedell,
  Phys. Rev. B {\bf 60}, 40 (1999).

\bibitem{jprev} for a review see J. Jakli\v c and P.  Prelov\v sek,
  Adv. Phys.  {\bf 49}, 1 (2000).


\bibitem{pine} J. Schmalian, D. Pines, and B. Stojkovi\'c, Phys. Rev.
  Lett. {\bf 80}, 3839 (1998); Phys. Rev. B {\bf 60}, 667 (1999).

\bibitem{dahm} T. Dahm, D. Manske, and L. Tewordt, Phys. Rev. B 
{\bf 60}, 14888 (1999).


\bibitem{prel} P.  Prelov\v sek, Z. Phys. B {\bf 103},
  363 (1997).

\bibitem{plak} N.M.  Plakida and V.S. Oudovenko, Phys. Rev.  B {\bf
    59}, 11949 (1999).

\bibitem{kane} {C.L. Kane, P.A. Lee, and N.  Read, Phys. Rev. B {\bf
    39}, 6880 (1989); G. Mart\'{\i}nez and P. Horsch, Phys. Rev. B {\bf
    44}, 317 (1991); A. Ram\v sak and P. Prelov\v sek,
 Phys. Rev. B {\bf 42}, 10415 (1990).}


\bibitem{lee} P.A. Lee, T.M. Rice, and P.W. Anderson, Phys. Rev. Lett.
  {\bf 31}, 462 (1973).

\bibitem{kamp} A.P. Kampf and J.R. Schrieffer, Phys. Rev. B {\bf 41},
  6399 (1990).

\bibitem{chub} A.V. Chubukov and D.K.  Morr, Phys. Rep. {\bf 288}, 355
  (1997).

\bibitem{mill} A.J. Millis and H. Monien,  Phys. Rev. B {\bf 61},
12496 (2000).

\bibitem{varm} C.M. Varma {\it et al.}, Phys. Rev. Lett. {\bf 63},
  1996 (1989).



\bibitem{bonca89} J. Bon\v ca, P. Prelov\v sek, and I. Sega,
Europhys. Lett. {\bf 10}, 87 (1989).  

\bibitem{putt} W.O. Putikka, M.U. Luchini, and R.R.P. Singh,
Phys. Rev. Lett. {\bf 81}, 2966 (1998).

\bibitem{tpr} T. Tohyama and S. Maekawa, J. Phys. Soc. Jpn.
{\bf 59}, 1760 (1990).  



\end{references}
\end{document}